\documentclass[%
 reprint,
 amsmath,amssymb,
 aps,
]{revtex4-2}
\usepackage[colorlinks=true, allcolors=blue]{hyperref}
\usepackage{float}
\usepackage{graphicx}
\usepackage{dcolumn}
\usepackage{bm}

\begin{document}
	
\preprint{APS/123-QED}

\title{Entangled-to-Packed Crossover in Nonlinear Extensional Rheology of Entangled Polymers}

\author{Yin Wang\textsuperscript{†}}
\author{Lin-Feng Wu\textsuperscript{†}}
\author{Yi-Bo Shao}
\author{Zhe Wang}
\thanks{† These authors contributed equally to this work}
\thanks{* Corresponding author}
 \email{zwang2017@mail.tsinghua.edu.cn}
 
\affiliation{%
 Department of Engineering Physics and Key Laboratory of Particle and Radiation Imaging (Tsinghua University) of Ministry of Education, Tsinghua University, Beijing 100084, China 
}%

\date{\today}

\begin{abstract}
Significant challenges exist in the nonlinear extensional rheology of entangled polymers. With simulations, we show that the key to understanding this problem is to recognize the existence and importance of a strain-induced crossover from the entangled state to a packed state. This crossover, following the saturation of primitive chain stretch, takes place with massive release of entanglements via convective flow and progressive chain alignment. After the crossover, the disentangled, fully-aligned chain segments pack similarly to the random packing of rods. Meanwhile, the system enters the steady state. In this state, the tube model—built on the concept of entanglement—fails, while the stress can be quantitatively calculated by combining the intra-chain conformational contribution and a frictional contribution from the inter-chain separation along flow.
\end{abstract}

\maketitle



Entanglements profoundly influence the rheology of polymers \cite{doi1988theory, doi1978dynamics-2, doi1978dynamics-3}. By conceptualizing discrete entanglements as a continuous “tube”, the tube model significantly simplifies the many-body interaction among chains~\cite{doi1988theory, schleger1998clear}. With refinements such as constraint release \cite{klein1978onset, marrucci1985relaxation}, and primitive length fluctuation \cite{doi1983explanation, milner1998reptation}, the tube model successfully describes the linear rheology of polymers, enabling a universal description of rheological behaviors across diverse systems \cite{MeadLarson-11, LikhtmanMcleish-9, GrahamLikhtman-10, van2002evaluation}. However, for the nonlinear extensional rheology, the validity of the tube model has been widely questioned in the past two decades \cite{wang2018nonlinear, wang2017fingerprinting, Likhtman-46, AlvarezHuang-15, dealy2018structure}. For example, the aforesaid universality is found to be invalid in nonlinear extensional flows, as suggested by the observation that solutions and melts with the same number of entanglements per chain $ Z $ exhibit opposite trends in the steady-state extensional viscosity as a function of strain rate \cite{AlvarezHuang-15,dealy2018structure, MatsumiyaWatanabe-12, BachAlmdal-13, BhattacharjeeOberhauser-14, BhattacharjeeNguyen-17, andre2021investigating}. To meet this crisis, several mechanisms have been proposed, such as inter-chain pressure \cite{MarrucciIanniruberto-18, WagnerKheirandish-20, NarimissaHuang-19}, nematic interaction \cite{DoiWatanabe-21, CaoLikhtman-22, ParkIanniruberto-23}, and friction reduction \cite{IannirubertoBrasiello-24, YaoitaIsaki-25, MasubuchiYaoita-26, IannirubertoMarrucci-27, BobbiliMilner-31,matsumiya2018nonlinear}. Even with these effects, tube model’s predictions still non-negligibly deviate from experimental and simulation results at large strains, especially in the steady state \cite{CostanzoHuang-16, hannecart2022decoding}. These results call for a critical revision of the current theoretical picture for the nonlinear extensional flow of entangled polymers.

To tackle this problem, we examine the behaviors of entanglements during extension. Previous studies show that in nonlinear flows, $ Z $ may dramatically decrease with increasing strain  \cite{BaigMavrantzas-29, LiuXu-144, WangWang-55, IannirubertoMarrucci-30}. Although this mechanism has been incorporated into tube models \cite{IannirubertoMarrucci-30, Ianniruberto-61}, entanglements’ microscopic dynamics remain poorly understood, thereby impeding understanding the elusive phenomena observed in nonlinear extensional flows. With molecular dynamics (MD) simulations \cite{KremerGrest-33}, we address the evolution of entanglements during the extensional flow, and its relation to the chain conformation, the inter-chain configuration, and most importantly, the rheology. We reveal an entangled-to-packed crossover in inter-chain configuration and interaction as the strain surpasses a critical value, corresponding to a fundamental change in system’s rheological response: At small strains, the system is well entangled, and its response can be quantified by the tube model. During the crossover, chains massively disentangle by convection, accompanied by pronounced chain alignment. After the crossover, the disentangled, aligned chain segments pack similarly to the random packing of rods. Meanwhile, the system is in the steady state. In this state, the tube model fails due to the radical disentanglement, while the rheological response can be described by combining the effects of the intra-chain conformation and the inter-chain friction.

We simulate a series of entangled melts composed of $ N_{\rm c} $ bead-spring chains with length $ N $ using LAMMPS \cite{Plimpton-34}. All beads interact via the repulsive Lennard-Jones potential, and bonded beads are connected by the finite extensible nonlinear elastic (FENE) potential \cite{HsuKremer-28, AuhlEveraers-56}. Reduced units are adopted with energy $ \epsilon $, length $\sigma $, and time $ \tau = \sigma(m/\epsilon)^{1/2} $, where $ m $ is the bead mass. The simulation is performed at temperature $ T=1 $ and density $ \rho=0.85 $. Extension along the $ z $ axis is applied under constant Hencky strain rate $ \dot{\varepsilon}_{\rm H} $ \cite{Dobson-35, Hunt-36,NicholsonRutledge-176}. To vary the equilibrium entanglement length  $ N_{\rm e} $, we modulate the chain stiffness by adjusting the bending coefficient $k_{\text{bc}}$ \cite{MoreiraZhang-38, AnwarGraham-37}. Additional details are provided in the End Matter (EM). Z1+ algorithm \cite{KrögerDietz-39} is used to identify primitive paths (PP) of chains and entanglements. Strain rate is represented by Rouse Weissenberg number $ {Wi}_{\rm R} = \dot{\varepsilon}_{\rm H} \tau_{\rm R} $, where $ \tau_{\rm R} $ is the chain Rouse time. $ {Wi}_{\rm R} $ is kept at $ {Wi}_{\rm R} \gg 1 $ to ensure strong nonlinear response.

\begin{figure}[ht]
  \centering
  \includegraphics[width=\linewidth]{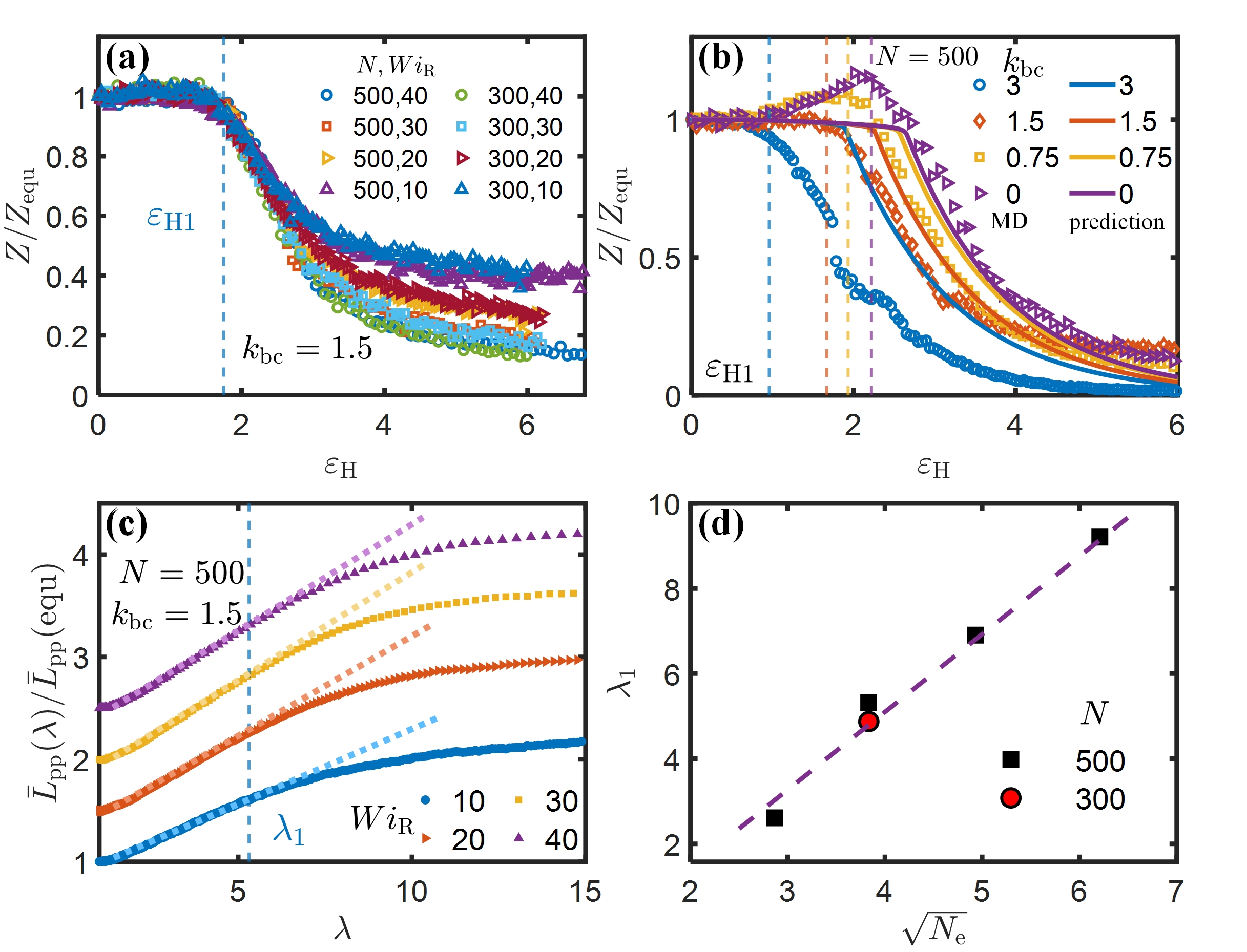}
  \caption{ $Z(\varepsilon_{\rm H})$, normalized by equilibrium value $Z_{\rm equ}$, for the conditions of (a) $k_{\text{bc}}=1.5$, $N=300$ and $500$, ${Wi}_{\rm R}=10 - 40$, and (b) $N=500$, ${Wi}_{\rm R}=40$, $k_{\text{bc}}=0 - 3$. Symbols: MD results; lines: model~\cite{IannirubertoMarrucci-30}  predictions (with CCR parameter $ \beta=0.8 $). (c) PP length growth $\bar{L}_{\rm pp}(\lambda)$ normalized by equilibrium value $\bar{L}_{\rm pp}(\rm equ)$ for the sample of $N=500$ and $k_{\text{bc}}=1.5$ at ${Wi}_{\rm R}=10 - 40$. Data are vertically shifted for clarity. In (a) – (c), the positions of $ \varepsilon_{\rm H1} $ or $ \lambda_1 $ are denoted by vertical dashed lines. (d) The critical stretch ratio $ \lambda_1 $ versus $ \sqrt{N_{\rm e}} $.}
  \label{fig:1}
\end{figure}

Figure~\ref{fig:1}(a) shows the evolutions of $ Z $ during extension with a fixed $ N_{\rm e} $ and various $ {Wi}_{\rm R} $ and $ N $. In all cases, $ Z $ exhibits an initial plateau, and starts to disentangle as the Hencky strain $ \varepsilon_{\rm H} $ surpasses a critical value $ \varepsilon_{\rm H1} $. Seen from Fig.~\ref{fig:1}(a), $ \varepsilon_{\rm H1} $ is independent on $ N $ and $ {Wi}_{\rm R} $. Figure~\ref{fig:1}(b) examines the evolutions of $ Z $ with different $ N_{\rm e} $, revealing a strong increase of $ \varepsilon_{\rm H1} $ on $ N_e $. To clarify this origin of the disentanglement, we check the mean PP length $\bar{L}_{\rm pp}$ as a function of the stretch ratio $ \lambda=e^{\varepsilon_{\rm H}}$, and show the results of a sample in Fig.~\ref{fig:1}(c). At small strains, $\bar{L}_{\rm pp}$ linearly grows with $ \lambda $, indicating an affine deformation. As $ \lambda $ surpasses $\lambda_1=e^{\varepsilon_{\rm H1}}$, the extension of $\bar{L}_{\rm pp}$ deviates from linearity and starts to saturate. The saturation of $\bar{L}_{\rm pp}$ signifies the breakdown of affine deformation, suggesting that the further extension involves significant sliding between chains, which leads to the release of entanglements. The sliding between chains has been attributed to the imbalance between inter-chain grip force and retraction force~\cite{SchweizerXie-62, XieSchweizer-63, WangRavindranath-57}. Considering that the maximum stretch ratio of PP is proportional to $N_{\rm e}^{1/2}$ \cite{O'ConnorAlvarez-53}, we expect a relation of $\lambda_{1}\sim N_{\rm e}^{1/2}$. This relation is confirmed in Fig.~\ref{fig:1}(d) with different samples. By employing the concept of convective constraint release (CCR)~\cite{marrucci1996dynamics,likhtman2000microscopic, read2004convective}, the loss of entanglement is modeled~\cite{IannirubertoMarrucci-30}, and the modeling results are also plotted in Fig.~\ref{fig:1}(b). It is seen that the CCR model qualitatively captures the features of the decay of $ Z $ observed in our MD simulation. The importance of convection in entanglement loss will be further clarified in following parts.

\textbf{\begin{figure}[ht]
		\centering
		\includegraphics[width=\linewidth]{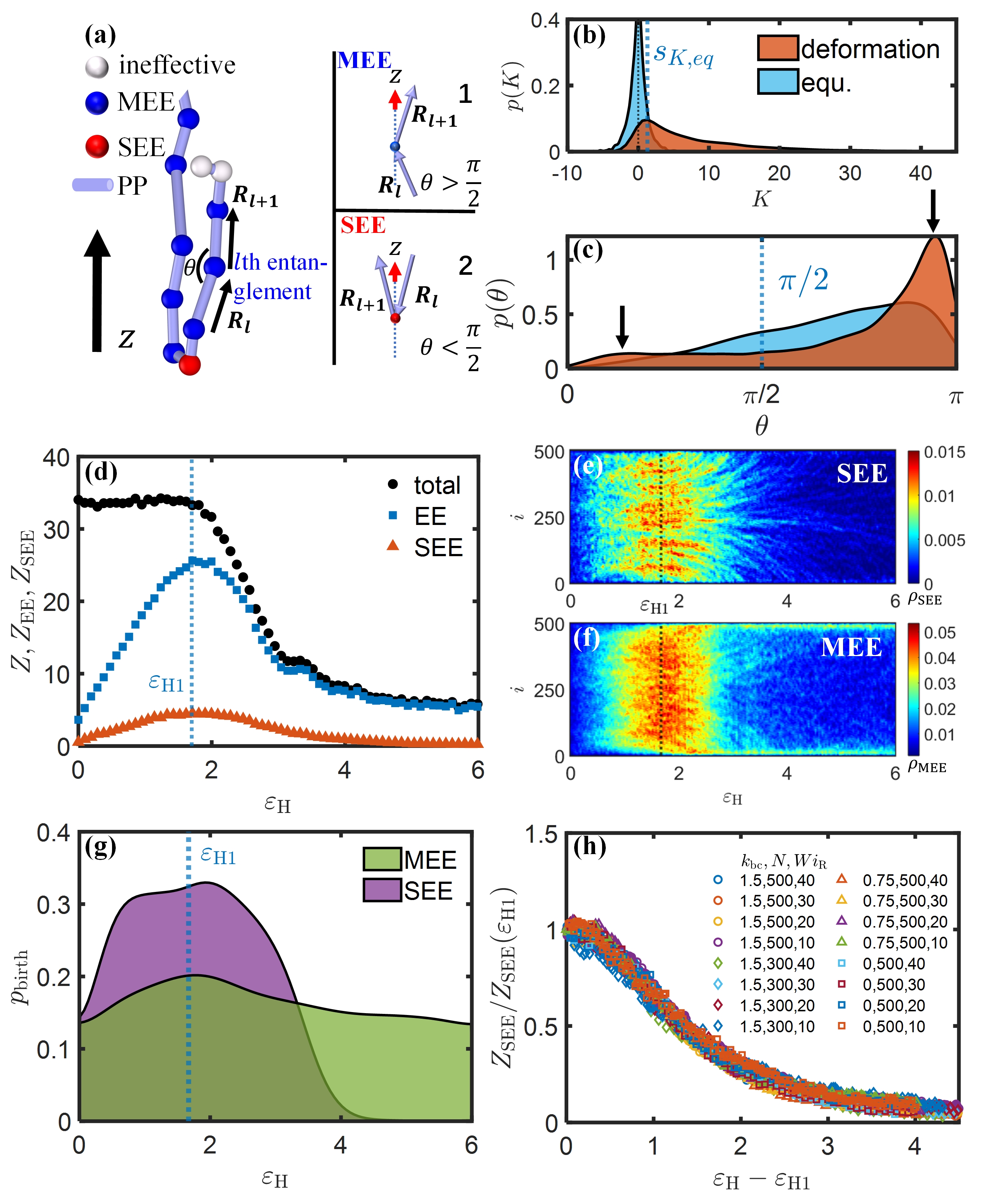}
		\caption{ Properties of EE. (a) Illustration of SEE, MEE and ineffective entanglement in a segment of PP stretched to $ \varepsilon_{\rm H1} $. (b) and (c) respectively give the distributions of $ K $ and $ \theta $ at equilibrium and at $ \varepsilon_{\rm H1} $. (d) Evolutions of total number of entanglements ($ Z $), EE ($ Z_{\rm EE} $), and SEE ($ Z_{\rm SEE} $) per chain during extension. (e) and (f) respectively give the density of entanglements $\rho(i, \varepsilon_{\rm H})$ along chain’s monomer coordinate $i$ as a function of  $ \varepsilon_{\rm H} $ for SEE and MEE. (g) Comparison between the distributions of birth time of SEE and MEE during extension. Data in (a) – (g) are measured with the sample of $N=500$ and $ k_{\rm bc}=1.5 $ at $ Wi_{\rm R}=40 $. (h) Decay of SEE at $ \varepsilon_{\rm H}\geq\varepsilon_{\rm H1} $ as a function of strain under various conditions.}
		\label{fig:2}
\end{figure}}

To further investigate the stretch-induced disentanglement, we focus on local structure and motion of entanglements during extension. It should be noted that not all entanglements found by PP identification methods, such as the Z1+ algorithm, effectively contribute to the stress in deformed polymers \cite{WuMao-40, RuanLu-54, HsuKremer-41, HsuKremer-42}. To account for this effect, we propose the concept of effective entanglement (EE) recently \cite{WuMao-40}. For a stretched chain, whether the $ l $th entanglement is an EE depends on the configuration of its two associated strands, reflected by a parameter $ K $:
\begin{equation}
	K(l) = \frac{R_{l,z}^2 - R_{l,x}^2}{N_{s,l}} + \frac{R_{l+1,z}^2 - R_{l+1,x}^2}{N_{s,l+1}},
	\label{eq:1}
\end{equation}
where \( R_{l,x/z} \) is the $x/z$ component of the end-to-end vector \( \textit{\textbf{R}}_l \) of the $l$th strand in a chain, and \( N_{s,l} \) is the number of monomers in the $l$th strand, as illustrated in Fig.~\ref{fig:2}(a). Seen from Fig.~\ref{fig:2}(b), as stretch is applied, the distribution of $ K $ deviates from the equilibrium form and shifts to larger values. An entanglement is identified as an EE if its $ K $ is larger than the standard deviation \(s_{K,\text{eq}} \) of the equilibrium $ K $ distribution \cite{WuMao-40}, which suggests that the conformation of its two strands effectively align with the flow direction and contribute to the conformational tensile stress $ \sigma_{\rm t}\sim \sum_l (R_{l,z}^2 - R_{l,x}^2)/{N_{s,l}} $.  


Entanglements with sharp angles have been discussed in literature, with observations highlighting their roles in the mechanical response of stretched polymers \cite{HsuKremer-41, HsuKremer-42, ZhengTsige-43, MoghadamSahaDalal-44, SmithChu-45}. For an EE, its angle $ \theta $ is given by $ \cos \theta = -\hat{\bm{R}}_l \cdot \hat{\bm{R}}_{l+1} $. According to $ \theta $, we classify all EEs into two types: sharp EE (SEE) with $ \theta \leq \pi/2 $, and mild EE (MEE) with $ \theta \geq \pi/2 $. In Fig.~\ref{fig:2}(a), we mark SEEs and MEEs in a PP of a stretched sample. The distributions of $ \theta $ ($ p(\theta) $) before and after a stretch to $ \varepsilon_{\rm H1} $ are shown in Fig.~\ref{fig:2}(c). The stretch causes $ p(\theta) $ to change from a unimodal form to a bimodal form, whose two peaks respectively correspond to SEE and MEE. Figure~\ref{fig:2}(d) shows the evolutions of total entanglements ($ Z $), EEs ($ Z_{\rm EE} $), and SEEs ($ Z_{\rm SEE} $) per chain of a stretched sample. Both $ Z_{\rm EE} $ and $ Z_{\rm SEE} $ increase with $ \varepsilon_{\rm H} $ in the beginning of extension due to chain orientation and stretching, and start to decrease at $ \varepsilon_{\rm H1} $. At  $ \varepsilon_{\rm H}>\varepsilon_{\rm H1} $, most entanglements become effective due to the strong alignment of chains. To examine the emergence and motion of EE, we calculate the local density of EE $\rho(i, \varepsilon_{\rm H})$ along the chain monomer coordinate $i$ ($i$ represents the $i$th monomer of a chain) as a function of $ \varepsilon_{\rm H} $. Figure~\ref{fig:2}(e) and (f) show the results of $\rho(i, \varepsilon_{\rm H})$ for SEE and MEE, respectively. Using the local maxima of $\rho(i, \varepsilon_{\rm H})$ to pinpoint the birth events of EEs, we obtain the distribution of birth event as a function of $ \varepsilon_{\rm H} $ for SEE and MEE, and show the results in Fig.~\ref{fig:2}(g). At $ \varepsilon_{\rm H}>\varepsilon_{\rm H1} $, the birth of SEE becomes rarer and rarer as $ \varepsilon_{\rm H} $ increases. Due to the acute feature, an SEE typically appears at the corner of a Z-fold PP, as illustrated in Fig.~\ref{fig:2}(a). Thus, the creation of an SEE requires a coiled conformation of the fragment of PP, which is increasingly suppressed as the chain alignment enhances. For an MEE, in contrast, the end-to-end vectors of its two associated strands are oriented to similar directions that are nearly parallel to the flow, as illustrated in Fig.~\ref{fig:2}(a). Consequently, MEEs can be generated even when chains are highly aligned. As seen in Fig.~\ref{fig:2}(g), MEE maintains stable formation rate at $ \varepsilon_{\rm H}>\varepsilon_{\rm H1} $. Additionally, at $ \varepsilon_{\rm H}>\varepsilon_{\rm H1} $, most MEEs frequently appear and release at chain ends due to fluctuations, as shown in Fig.~\ref{fig:2}(f). Therefore, MEEs form a “dynamic background” contribution to the total number of entanglements at large strains. On the other hand, as shown in Fig.~\ref{fig:2}(e), SEEs show clear outward-sliding trajectories along the PP as strain proceeds, revealing the convective origin of their release.

As discussed above, beyond $ \varepsilon_{\rm H1} $, SEEs are of significant geometrical importance and are rare to be regenerated, thus serving as an indicator for probing the disentangling process. In Fig.~\ref{fig:2}(h), we plot the decay of $ Z_{\rm SEE}(\varepsilon_{\rm H}) $ at $ \varepsilon_{\rm H}>\varepsilon_{\rm H1} $, normalized at $ \varepsilon_{\rm H}=\varepsilon_{\rm H1} $, for different samples and flow rates. It is seen that all data collapse onto a master curve, suggesting that the disentangling process takes place through the convection of flow. A model for the convective decay of $ Z_{\rm SEE}(\varepsilon_{\rm H}) $ is provided in the EM.

\textbf{\begin{figure}[ht]
		\centering
		\includegraphics[width=\linewidth]{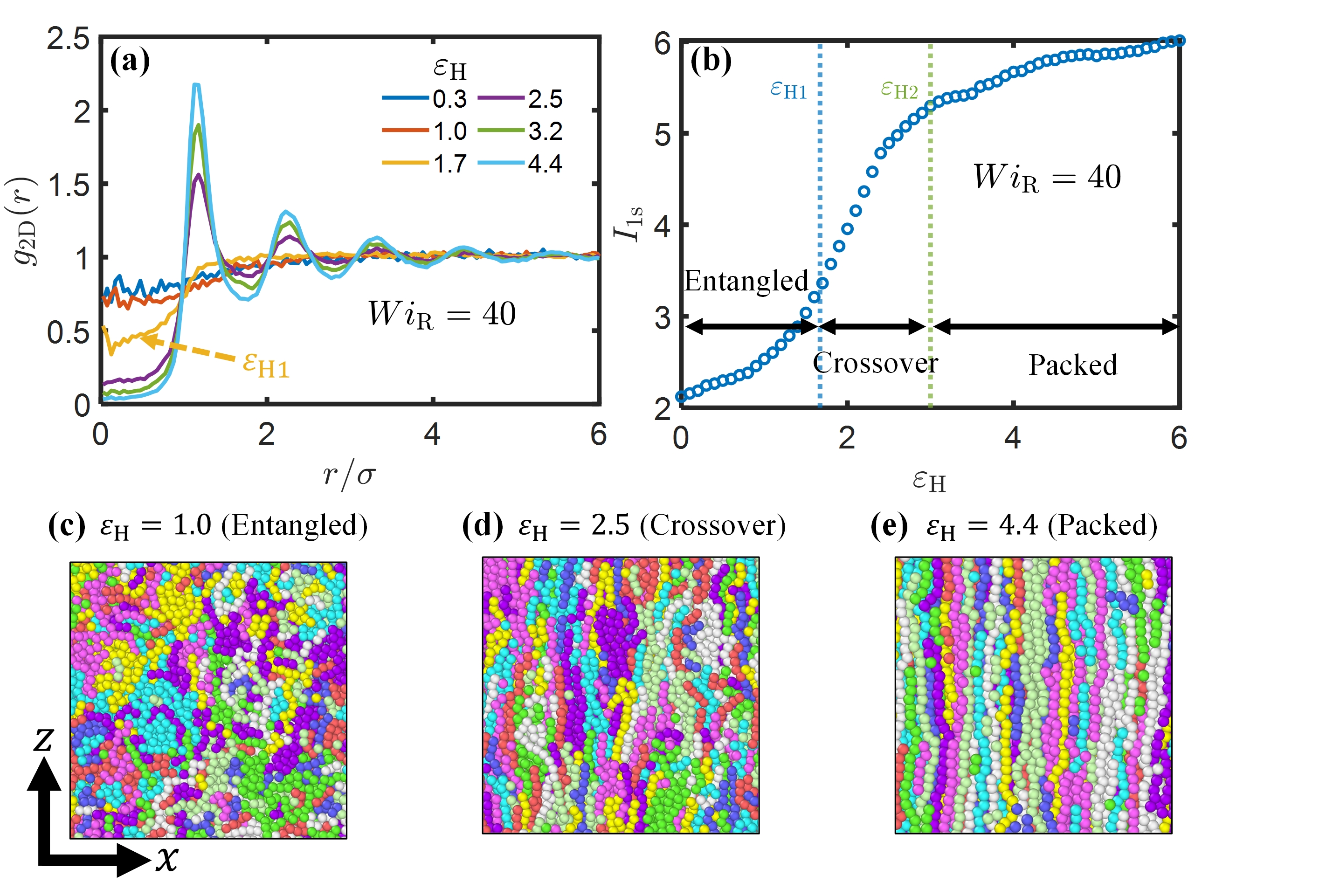}
		\caption{ Structural evolution of the sample of $N=500$ and $k_{\rm bc}=1.5$ at $ Wi_{\rm R}=40 $. (a) Evolution of $g_{\rm 2D}(r)$ during extension. (b) $ I_{\rm 1s} $ as a function of $ \varepsilon_{\rm H} $. Two critical strains, $ \varepsilon_{\rm H1} $ and $ \varepsilon_{\rm H2} $, are denoted by vertical dashed lines. (c) – (e) respectively show the snapshots of chains at $ \varepsilon_{\rm H}=1.0 $, 2.5, and 4.4.}
		\label{fig:3}
\end{figure}}

With progressive stretching, the population of SEEs diminishes to nearly zero, meanwhile the chain conformation gradually achieves full-chain alignment. To examine the influence of these phenomena on the inter-chain configuration, we calculate the 2-dimensional (2D) pair distribution function $g_{\rm 2D}(r)$ of chains. To be specific, we first partition the system into slices along flow direction with certain thickness. Then, for each slice, we project the centers of mass of chain segments within this slice onto the $ x-y $ plane. With these projected points, $g_{\rm 2D}(r)$ can be calculated (see EM). Figure~\ref{fig:3}(a) shows the evolution of $g_{\rm 2D}(r)$ of a sample as $ \varepsilon_{\rm H} $ increases with $ Wi_{\rm R}=40 $. At $ \varepsilon_{\rm H}>\varepsilon_{\rm H1} $, $g_{\rm 2D}(r)$ gradually develops a pronounced peak, corresponding to a well-defined first coordination shell. With $g_{\rm 2D}(r)$, we calculate the average chain number in the first shell $ I_{\rm 1s} $. As shown in Fig.~\ref{fig:3}(b), $ I_{\rm 1s} $ rapidly grows with $ \varepsilon_{\rm H} $ when $ \varepsilon_{\rm H} > \varepsilon_{\rm H1} $, and starts to saturate to a plateau value close to 6 when $ \varepsilon_{\rm H} $ exceeds another critical value $ \varepsilon_{\rm H2} $. The behavior of $g_{\rm 2D}(r)$ at $ \varepsilon_{\rm H}>\varepsilon_{\rm H2} $ exhibits typical features of 2D random packing \cite{mulero2008theory}. Figure~\ref{fig:3}(c)-(e) respectively display the chains of a sample stretched to $ \varepsilon_{\rm H}=1 $ ($ <\varepsilon_{\rm H1} $), 2.5 ($ \varepsilon_{\rm H1}<\varepsilon_{\rm H}<\varepsilon_{\rm H2} $) and 4.4 ($ >\varepsilon_{\rm H2} $) with $ Wi_{\rm R}=40 $, whose difference visualizes the entangled-to-packed crossover in the inter-chain configuration, as suggested by the behaviors of $g_{\rm 2D}(r)$. Seen from Fig.~\ref{fig:2}, a few MEEs still exist at $ \varepsilon_{\rm H}>\varepsilon_{\rm H2} $, while they are rapidly fluctuating and well aligned with the flow, thereby hardly affecting the packed structure. Note that, some studies have implied the emergence of the packed state at large strains but lack systematic demonstration and analysis \cite{O'ConnorAlvarez-53, IannirubertoMarrucci-27, lopez2017molecular, o2019stress}. Our work clearly shows the existence of such state, and how it is induced by the strong extensional flow.

\textbf{\begin{figure}[ht]
		\centering
		\includegraphics[width=\linewidth]{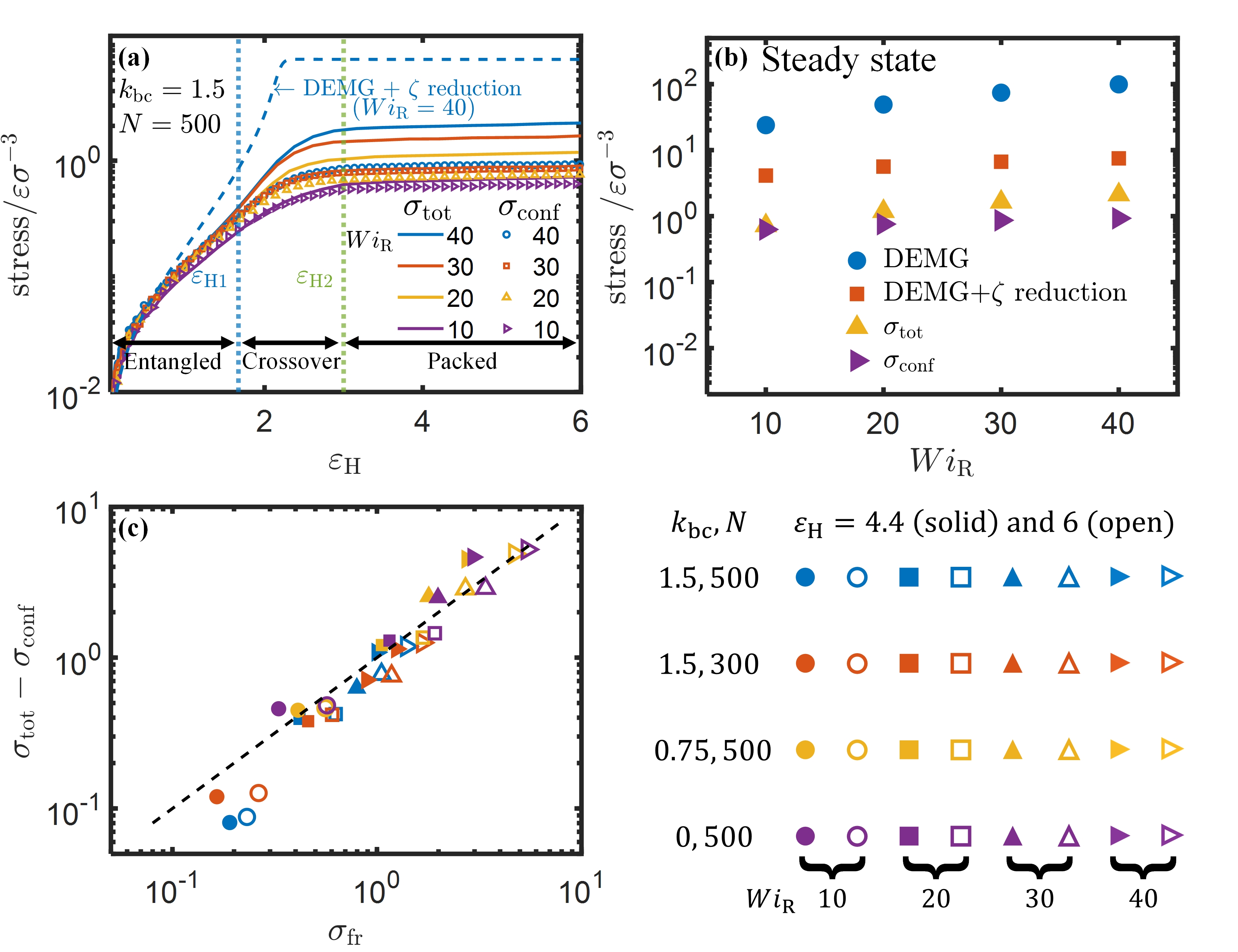}
		\caption{ Tensile stress. (a) Total stress $ \sigma_{\rm tot} $ and conformational stress $ \sigma_{\rm conf} $ as a function of $ \varepsilon_{\rm H}$ for the sample of $N=500$ and $k_{\rm bc}=1.5$ at $Wi_{\rm R}=10-40$. DEMG  prediction with friction reduction is also given (dashed line). (b) $ \sigma_{\rm tot} $, $ \sigma_{\rm conf} $, and DEMG predictions (with and without friction reduction) at steady states as a function of  $Wi_{\rm R}$ for the sample of $N=500$ and $k_{\rm bc}=1.5$. (c) Scatter plot between $ \Delta\sigma=\sigma_{\rm tot}-\sigma_{\rm conf} $ and frictional stress $ \sigma_{\rm fr} $ (eq.~\ref{eq:3}) for various steady-state conditions denoted in the right side of panel. The dashed line denotes $ \Delta\sigma=\sigma_{\rm fr} $.}
		\label{fig:4}
\end{figure}}

In the packed state, concepts built on entanglements, such as $N_{\rm e}$ and the reptation time $ \tau_{\rm d} $, lose their structural basis. This explains why the tube model cannot give a satisfactory description on the nonlinear extensional rheology. Figure~\ref{fig:4}(a) shows the stress-strain data of a sample at different $ Wi_{\rm R}$. The total tensile stress $ \sigma_{\rm tot} $, calculated by the Virial expression \cite{AnwarGraham-37, HsuKremer-28}, enters the steady state at $ \varepsilon_{\rm H}>\varepsilon_{\rm H2} $ where both the chain conformation and inter-chain configuration become steady. The result at $ Wi_{\rm R}=40 $ predicted by the DEMG model supplemented by FENE and friction reduction (see EM) \cite{DesaiLarson-59, PearsonHerbolzheimer-60} is also shown in Fig.~\ref{fig:4}(a). The theoretical result remarkably deviates from the MD result at $ \varepsilon_{\rm H}>\varepsilon_{\rm H1} $, suggesting the failure of the tube model. An alternative picture is, thus, in need. To seek this picture, we first calculate the conformational stress via the stress-optical rule (SOR) by \cite{LuapMüller-164, GaoWeiner-173, CaoLikhtman-49, WatanabeMatsumiya-143}:
\begin{equation}
	\sigma_{\rm conf} = \frac{
		\sum_{i=1}^{N_{\rm c}} \sum_{j=1}^{N-1} \left[ u_{ij}^z u_{ij}^z - \frac{1}{2} \left( u_{ij}^x u_{ij}^x + u_{ij}^y u_{ij}^y \right)\right] 
	}{\alpha V},
	\label{eq:2}
\end{equation}
where $u_{ij}^{\beta}$ is the $\beta$ component of the $ j $th bond vector in the $ i $th chain. The parameter $ \alpha $ is determined by equating $ \sigma_{\rm conf} $ to the Virial result $ \sigma_{\rm tot} $ in the low-strain regime where the SOR holds \cite{WatanabeMatsumiya-143, lopez2019local}. The results of $ \sigma_{\rm conf} $ are also given in Fig.~\ref{fig:4}(a). $ \sigma_{\rm conf} $ and $ \sigma_{\rm tot} $ exhibit nonnegligible difference at $ \varepsilon_{\rm H}>\varepsilon_{\rm H1} $, which is enhanced as $ Wi_{\rm R} $ increases. The systematic difference between $ \sigma_{\rm conf} $ and $ \sigma_{\rm tot} $ in the steady state, and that between MD results and DEMG predictions, are summarized in Fig.~\ref{fig:4}(b).

As shown in Fig.~\ref{fig:4}(b), at $ \varepsilon_{\rm H}>\varepsilon_{\rm H2} $,  $ \sigma_{\rm conf} $ alone markedly underestimates the total stress. The revised DEMG model accounts for this by including FENE stretching, but consequently overpredicts the steady-state stress even with the friction reduction~\cite{CostanzoHuang-16, DesaiLarson-59}. Moreover, although previous work has incorporated entanglement loss (Fig.~\ref{fig:1}(b)), the steady-state stress remains overestimated due to the retention of FENE~\cite{IannirubertoMarrucci-30, Ianniruberto-61}. Herein, we attribute the stress difference $ \Delta\sigma = \sigma_{\rm tot} - \sigma_{\rm conf} $ to the frictional stress $ \sigma_{\rm fr} $ that develops during the entangled-to-packed crossover, arising from the strain-induced inter-chain separation along flow. The frictional stress between aligned, unentangled chains is given by $ \sigma_{\rm fr,0}\sim \rho\zeta \dot{\varepsilon}_{\rm H} {\bar L}_{\rm z}^{2}/12 $~\cite{O'ConnorAlvarez-53, IannirubertoMarrucci-27}, where $ {\bar L}_{\rm z} $ is the average chain extension along flow and $ \zeta $ is the monomer friction coefficient (evaluated by a chain dragging method, detailed in the EM). Then, considering that the residual entanglements can suppress the inter-chain separation, the frictional stress can be written as
\begin{equation}
	\sigma_{\rm fr}=C\left(\frac{Z}{Z_{\rm equ}}\right) \frac{\rho\zeta \dot{\varepsilon}_{\rm H} {\bar L}_{\rm z}^{2}}{12},
	\label{eq:3}
\end{equation}
where $ C(\cdot) $ is a coefficient depending on the residual entanglements $ Z $ normalized by the equilibrium value $ Z_{\rm equ} $. In steady states, $ Z / Z_{\rm equ} $ is a small value, enabling an expansion form $ C\approx 1-fZ/Z_{\rm equ} $. We set $ f=1 $ to ensure that $ C\approx0 $ at $ Z = Z_{\rm equ} $, where the inter-chain separation along flow is completely suppressed by entangled network. Figure~\ref{fig:4}(c) demonstrates that this expression for $ \sigma_{\rm fr} $ reproduces $ \Delta\sigma $ across varying chain lengths, stiffnesses, and flow rates. In contrast, for entangled states ($ \varepsilon_{\rm H}<\varepsilon_{\rm H1} $, $ C\approx0 $), the entangled network hinders the inter-chain separation along flow, rendering this frictional stress (and $ \Delta\sigma $) negligible, as shown in Fig.~\ref{fig:4}(a).

According to the above analysis, friction’s role in rheology undergoes a transition due to the entangled-to-packed crossover. Near equilibrium, frictional effect is indirect, “hidden” within the conformational dynamics through controlling the conformational relaxation time \cite{doi1988theory}.  After the crossover, inter-chain separation along flow becomes important due to the exhaustive constraint release and chain alignment, leading to a direct stress contribution from the inter-chain friction. Though some studies have emphasized the role of inter-chain interactions in stress \cite{Likhtman-46, gao1994nature, RamírezSukumaran-167, GaoWeiner-172}, this insight has not been incorporated into subsequent tube models. Moreover, some studies imply the dominant role of inter-chain friction in the steady state by employing a picture of “rigid chains”, in which the stress is purely from the inter-chain dissipation \cite{O'ConnorAlvarez-53, IannirubertoMarrucci-27}. Nevertheless, for non-rigid aligned chains, such as our MD chains, their conformation is dynamically stabilized with intra-chain modes of stretching and retracting, through which the energy is repeatedly stored and released. Thus, the inter-rigid-chain dissipation alone cannot constitute the total stress. By combining both the intra-chain conformation and the inter-chain friction, we nicely quantify the steady-state response.

In summary, we reveal an entangled-to-packed crossover in inter-chain configuration in the nonlinear extension of entangled polymers. Following the saturation of primitive chain stretch, this crossover is accompanied by massive release of entanglements by convection, and finally leads to the steady state, in which the disentangled, aligned chains randomly pack. In the packed state, the tube model fails, while the rheology can be understood by decomposing the stress into a conformational part and a frictional contribution from inter-chain separation, with the latter being long neglected in most theoretical pictures. 

We are grateful to Prof. Kröger for helpful discussion and Dr. Yangyang Wang for previous collaboration. This research was supported by National Natural Science Foundation of China (no. 11975136).

\bibliographystyle{unsrt}
\bibliography{apssamp}

\clearpage
\begin{center}
	\textbf{\large End Matter}
\end{center}
\vspace{0.5em}
\appendix

\textit{Appendix A: Simulation} -- Simulations of entangled polymer melts are based on the standard Kremer–Grest model~\cite{HsuKremer-28}. Nonbonded monomers interact via the WCA potential \( U_{\mathrm{WCA}} = 4\epsilon \left[ \left( \frac{\sigma}{r} \right)^{12} - \left( \frac{\sigma}{r} \right)^6 + \frac{1}{4} \right] \) for \( r < 2^{1/6} \sigma \), and 0 otherwise. Consecutive beads with mean bond length \( b = 0.964 \) are connected by the FENE potential \( U_{\mathrm{FENE}} = -\frac{1}{2} k R_0^2 \ln \left[ 1 - \left( \frac{r}{R_0} \right)^2 \right] \) with \( k = 30\epsilon / \sigma^2 \)  and \( R_0 = 1.5\sigma \). Chain stiffness is introduced through \( U_{\mathrm{bend}} = k_{\rm bc} (1 - \cos \alpha ) \), where \( \alpha \) is the bond angle. We simulate \( N_{\rm c} = 250 \) or 200 chains of length \( N = 300 \) or 500 at a fixed monomer density \( \rho = N_{\rm c} N/V = 0.85\,\sigma^{-3} \). Equilibrated configurations are generated using a Monte Carlo bond-swap algorithm~\cite{AuhlEveraers-56} and validated against established benchmarks~\cite{HsuKremer-28}. For \( k_{\rm bc} = 0 \), 0.75, and 1.5, the entanglement lengths are \( N_{\rm e} \approx 60 \), 51 and 28, corresponding to entanglement times \( \tau_{\rm e} \approx 3290\tau \), \( 2800\tau \) and \( 1980\tau \), respectively~\cite{HsuKremer-28, MoreiraZhang-38, AnwarGraham-37, CaoLikhtman-49}. Rouse and reptation times are estimated as \( \tau_{\rm R} = \tau_{\rm e} \left( N/{N_{\rm e}} \right)^2 \) and \( \tau_{\rm d} = 3 \tau_{\rm e} \left(N/{N_{\rm e}} \right)^3 \), with the plateau modulus \( G_{\rm N}^0 = \frac{4}{5} \frac{\rho k_{\rm B} T}{N_{\rm e}} \), where \( k_{\rm B} \) is Boltzmann constant. Flow is imposed via the SLLOD equations of motion, combined with a Nosé–Hoover thermostat and the generalized Kraynik–Reinelt boundary condition~\cite{Dobson-35, Hunt-36, NicholsonRutledge-176}.

\textit{Appendix B: Convective SEE release} -- To model the release of SEE, we regard a Z-fold PP as being connected to an elastic plate that deforms affinely with \( \dot{\varepsilon}_{\mathrm{H}} \). SEEs are treated as rings pinned to the plate (Fig.~\ref{fig:5}(a)). As the plate elongates, rings slide along the PP and are released once reaching a chain end (Fig.~\ref{fig:5}(a)). Here we introduce a threshold strain \( \varepsilon_{\mathrm{HR}}\), in the sense that for \( \varepsilon_{\mathrm{H}} > \varepsilon_{\mathrm{HR}} \) the generation of new SEE is ignorable. \( \varepsilon_{\mathrm{HR}}\) is larger than \( \varepsilon_{\mathrm{H1}}\). We then consider the convective release of SEE at \( \varepsilon_{\mathrm{H}} > \varepsilon_{\mathrm{HR}} \). Let \( x(t^{\prime})  = \frac{Z_{\mathrm{SEE}}(\varepsilon_{\mathrm{HR}} + \dot{\varepsilon}_{\mathrm{H}} t^{\prime})}{Z_{\mathrm{SEE}}(\varepsilon_{\mathrm{HR}})} \) denote the fraction of remaining SEE at time $ t^{\prime} $ (the origin of time is set as the moment the PP elongates to \( \varepsilon_{\mathrm{H}} = \varepsilon_{\mathrm{HR}} \)). Considering a SEE at the contour coordinate $ s=x(t^{\prime})L_{\rm sc} $ at $ t=0 $, where $ L_{\rm sc} $ is the half length of PP, and the origin of $ s $ is at the center of PP. At $ t = t^{\prime} $, it just reaches the end of PP ($ s = L_{\rm sc} $). It is easy to see that SEEs with $ s<x(t^{'})L_{\rm sc} $ at $ t=0 $ will survive at $ t = t^{\prime} $, while those with $ s>x(t^{\prime})L_{\rm sc} $ at $ t=0 $ will release at $ t = t^{\prime} $. From $ t=0 $ to $ t^{\prime} $, the plate elongates with the ratio $ \lambda = \exp(\dot{\varepsilon}_{\mathrm{H}} t^{\prime}) $. Then, it is found that $ \lambda = \exp(\dot{\varepsilon}_{\mathrm{H}} t^{\prime})=\frac{L_{\rm sc}}{x(t^{\prime})L_{\rm sc}}=\frac{1}{x(t^{\prime})} $, which gives $ x(t) = \exp({-\dot{\varepsilon}_{\mathrm{H}} t}) $. Note that this result is based on the assumption that PP length is fixed. In fact, even at \( \varepsilon_{\mathrm{H}} > \varepsilon_{\mathrm{HR}} \), PP length still mildly increases with strain. To correct this problem, we can incorporate the stretch of PP \( \lambda_{\mathrm{pp}}(t) = \overline{L}_{\mathrm{pp}}(t) / \overline{L}_{\mathrm{pp}}(0) \), which results in:

\renewcommand{\theequation}{B\arabic{equation}} 
\setcounter{equation}{0} 
\begin{equation}
	x(t) = \frac{\lambda_{\mathrm{pp}}(\varepsilon_{\mathrm{HR}} + \dot{\varepsilon}_{\mathrm{H}} t)}{\lambda_{\mathrm{pp}}(\varepsilon_{\mathrm{HR}})} \cdot \frac{1}{\exp(\dot{\varepsilon}_{\mathrm{H}} t)},
	\label{eq:B1}
\end{equation}
$ \lambda_{\rm pp} $ can be obtained from MD data. Using \( \varepsilon_{\mathrm{HR}}=2.2\), Fig.~\ref{fig:5}(b) compares the MD results of the sample with \( k_{\mathrm{bc}} = 1.5 \), \( N = 500 \) and \( \mathrm{Wi}_{\mathrm{R}} = 40 \) with predictions of eq.~\ref{eq:B1} for four different $ Wi_{\rm R} $ values. All model curves collapse onto a master curve and agree with the simulation results, demonstrating that SEE release is governed by convection.

\textbf{\begin{figure}[ht]
		\centering
		\includegraphics[width=\linewidth]{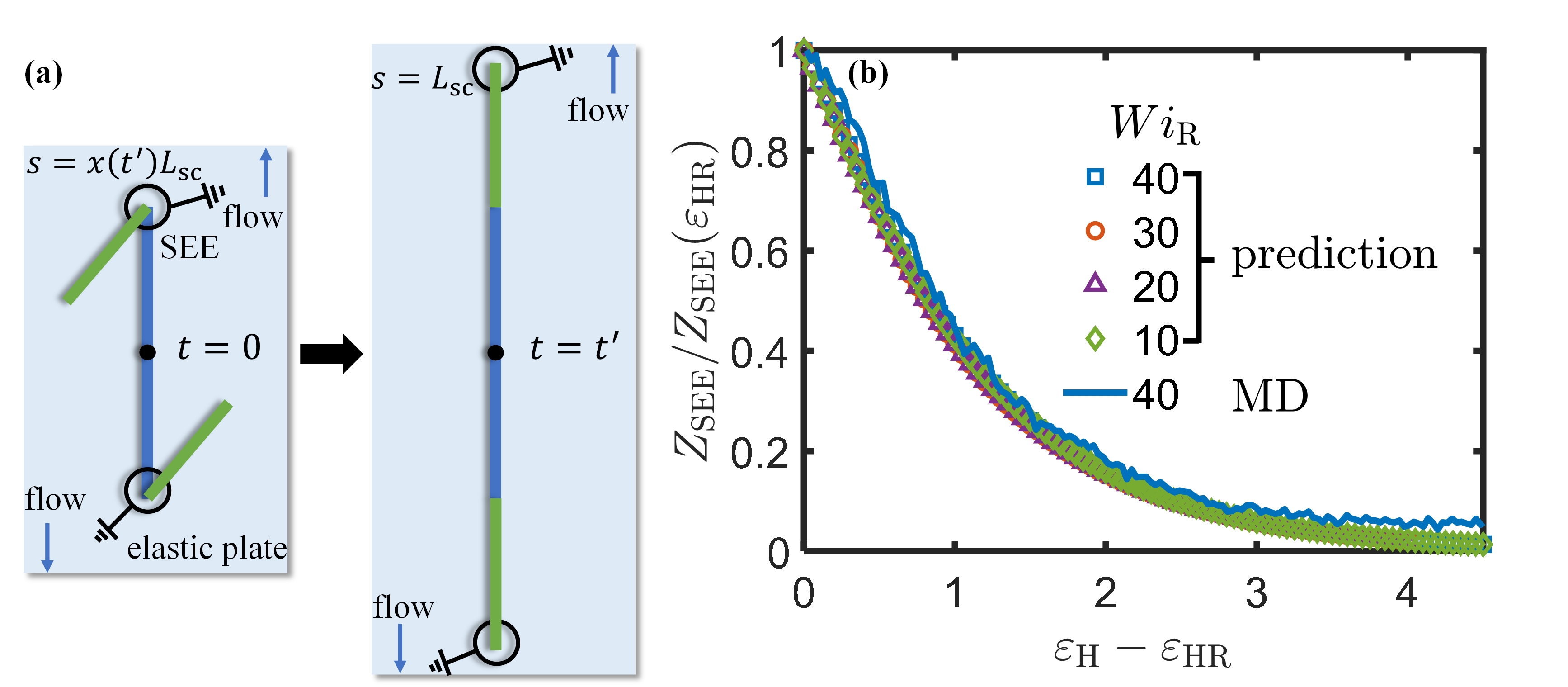}
		\caption{ Geometric mechanism of SEE release for a sample with \( N = 500 \) and \( k_{\mathrm{bc}} = 1.5 \). (a) Illustration of SEE sliding and release along the primitive path during affine deformation of a Z-fold chain. (b) Remaining SEE fraction for \( \varepsilon_{\mathrm{HR}} = 2.2 \): MD results compared to analytic predictions (\( \mathrm{Wi}_{\mathrm{R}} = 10\text{--}40 \)).}
		\label{fig:5}
\end{figure}}

\textit{Appendix C: }\( g_{\mathrm{2D}}(\boldsymbol{r}) \) -- We slice the simulation box along $ z $-axis with each slice having a thickness of \( \sqrt{N_{\mathrm{e}}}\, b \). For each slice, the centers of mass of chain segments contained in this slice are projected onto the \( x \)-\( y \) plane (Fig. \ref{fig:6}), resulting in the projected coordinate sets \( \{  \boldsymbol{r}_l  \} \). Then, its \( g_{\mathrm{2D}}(\boldsymbol{r}) \) is given by

\renewcommand{\theequation}{C\arabic{equation}} 
\setcounter{equation}{0} 
\begin{equation}
	\rho_{\mathrm{2D}} g_{\mathrm{2D}}(\boldsymbol{r}) = \frac{1}{N_{\mathrm{2D}}} \sum_{l=1}^{N_{\mathrm{2D}}} \sum_{\substack{l' \neq l}}^{N_{\mathrm{2D}}} \left\langle \delta\!\left[ \boldsymbol{r} + \boldsymbol{r}_l - \boldsymbol{r}_{l'} \right] \right\rangle.
	\label{eq:C1}
\end{equation}
where \( N_{\mathrm{2D}} \) is the number of projected points and \( \rho_{\mathrm{2D}} \) is the 2D density of projected points for the slice. The \( g_{\mathrm{2D}}(\boldsymbol{r}) \) of the system is found by averaging the results from all slices. With \( g_{\mathrm{2D}}(\boldsymbol{r}) \), $ I_{\rm 1s} $ is found by $I_{1s} = 2\pi\rho_{\rm 2D} \int_{0}^{r_{\min}} g_{\rm 2D}(r) r \, dr$, where $ r_{\min} $ is the position of \( g_{\mathrm{2D}}(\boldsymbol{r}) \)'s first minimum.

\textbf{\begin{figure}[ht]
		\centering
		\includegraphics[width=\linewidth]{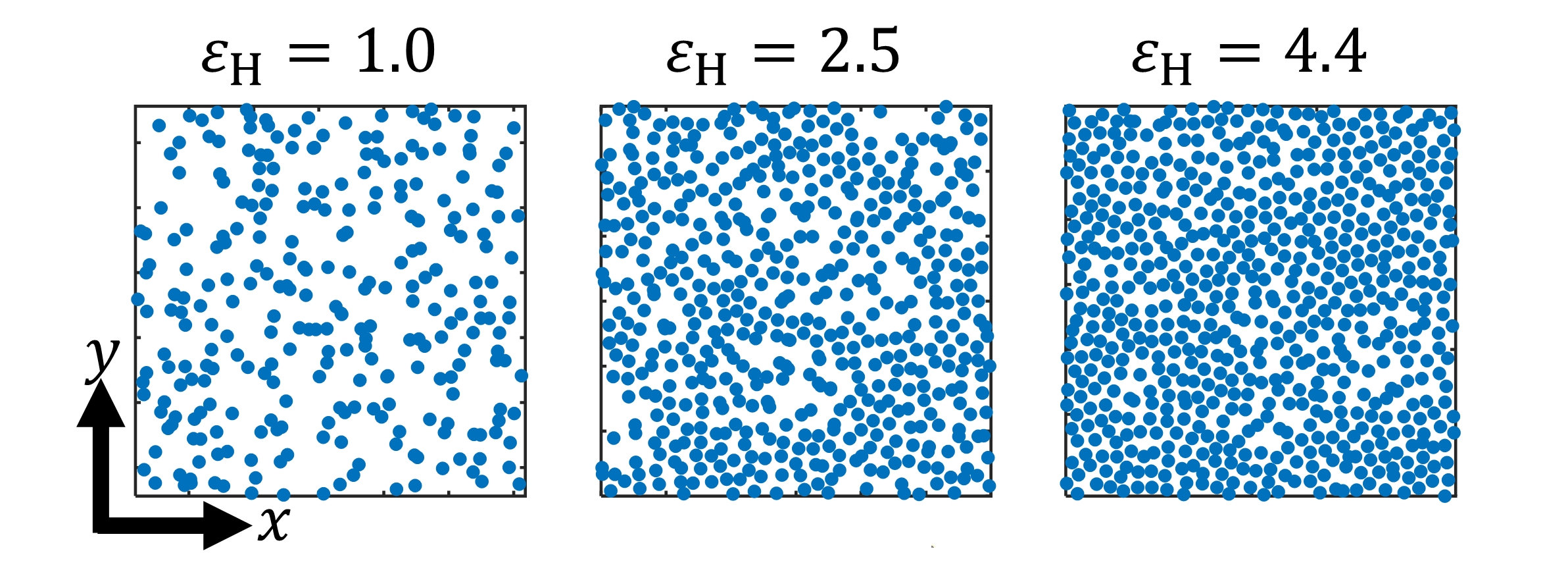}
		\caption{Snapshot of $ x-y $ plane and the corresponding projected centers-of-mass \( \{  \boldsymbol{r}_l  \} \) of chain segments, for a sample with \( N = 500 \), \( k_{\mathrm{bc}} = 1.5 \) and \( {Wi}_{\mathrm{R}} = 40 \) at \( \varepsilon_{\mathrm{H}} = 1.0 \), \( 2.5 \) and \( 4.4 \).}
		\label{fig:6}
\end{figure}}

\textit{Appendix D: DEMG model} -- The differential form of the DEMG model without friction reduction is given by~\cite{DesaiLarson-59, PearsonHerbolzheimer-60}:

\renewcommand{\theequation}{D\arabic{equation}} 
\setcounter{equation}{0} 
\begin{equation}
	\overset{\triangledown}{\boldsymbol{S}} = -2(\boldsymbol{\kappa} : \boldsymbol{S}) \boldsymbol{S} - \frac{1}{\lambda^{2} \tau_{\mathrm{d}}} \left( \boldsymbol{S} - \frac{\boldsymbol{\delta}}{3} \right),
	\label{eq:D1}
\end{equation}
\begin{equation}
	\frac{d\lambda}{dt} = \lambda (\boldsymbol{\kappa} : \boldsymbol{S}) - \frac{k_{\rm s} (\lambda - 1)}{\tau_{\mathrm{R}}},
	\label{eq:D2}
\end{equation}
\begin{equation}
	\boldsymbol{\sigma} = 3 G_{\rm N}^{0} k_{\rm s} \lambda^{2} \boldsymbol{S}.
	\label{eq:D3}
\end{equation}
Here, $\boldsymbol{S}$ and $\lambda$ denote the tube orientation tensor and stretch ratio, respectively; $\overset{\triangledown}{\boldsymbol{S}}$ represents the upper-convected derivative, $\boldsymbol{\kappa}$ is the transpose of the velocity gradient tensor, and $\boldsymbol{\delta}$ is the unit tensor. Model parameters $\tau_{\mathrm{R}}$, $\tau_{\mathrm{d}}$ and $G_{\rm N}^{0}$ are matched to our simulated systems. The nonlinear spring coefficient $k_{\rm s}$ is given by the inverse Langevin function approximation, $k_{\rm s} = \frac{(3\lambda_{\mathrm{max}}^{2} - \lambda^{2})/(\lambda_{\mathrm{max}}^{2} - \lambda^{2})}{(3\lambda_{\mathrm{max}}^{2} - 1)/(\lambda_{\mathrm{max}}^{2} - 1)}$, where the $\lambda_{\mathrm{max}} = \frac{N_{\mathrm{e}} b}{\sqrt{N_{\mathrm{e}}}\, b} = \sqrt{N_{\mathrm{e}}}$ is the maximum stretch ratio of chain.

To incorporate flow-induced friction reduction, the relaxation times reduce according to  $\tau_{\mathrm{R}} = \tau_{\mathrm{R}}^{0} \zeta(F_{\mathrm{so}})$ and $\tau_{\mathrm{d}} = \tau_{\mathrm{d}}^{0} \zeta(F_{\mathrm{so}})$, where $\tau_{\mathrm{R}}^{0}$ and $\tau_{\mathrm{d}}^{0}$ are equilibrium values, and the friction reduction factor is $\zeta(F_{\mathrm{so}}) = \left( 1 - \frac{\lambda^{4}}{\lambda_{\mathrm{max}}^{4}} (\boldsymbol{S} : \boldsymbol{S}) \right)^{2}$~\cite{DesaiLarson-59}.

\textit{Appendix E: Monomer friction coefficient} -- For undeformed entangled systems, monomer friction can be found by the fluctuation–dissipation theorem, where the diffusion coefficient is measured and converted to \( \zeta \) using the Einstein relation~\cite{BobbiliMilner-31, IannirubertoMarrucci-27}. In steady states, chains are disentangled and aligned. In this case, inter-chain separation along flow becomes significant. Thus, we can directly determine \( \zeta \) via a chain dragging method. To find \( \zeta \) at a strain $\varepsilon_{\rm H}$ ($ \varepsilon_{\rm H}>\varepsilon_{\rm H2} $), we drag a test chain with length \( N_{\mathrm{t}} \) along $ z $ axis in the stretched environment composed of stretched surrounding chains. To generate these surrounding chains, we stretch the sample to $\varepsilon_{\rm H}$, and then fix the end beads of all chains to keep their orientations and stretching at $\varepsilon_{\rm H}$. The test chain is subjected to a pair of balanced forces applied at its two ends, respectively along $ z $ and $ -z $ directions. The magnitude of this balanced force is adjusted to keep the stretch ratio of the test chain consistent with those of the chain segments of the same length $ N_{\rm t} $ in the sample stretched to $\varepsilon_{\rm H}$. Uniform drag forces \( F \) are applied to every bead of the test chain along the flow direction with the same magnitude. The simulation is thermostatted using the Nosé–Hoover thermostat. As shown in Fig.~\ref{fig:7}(a) for the case of \( N_{\mathrm{t}} = 500 \) and \( \varepsilon_{\mathrm{H}} = 6 \), the center-of-mass velocity $v$ of test chain along flow direction increases linearly with the applied drag force $F$, confirming the validity of the linear friction law. Additionally, we find that the extracted monomer friction coefficient \( \zeta \) is not sensitive to \( N_{\mathrm{t}}\) for \( N_{\mathrm{t}} \geq N_{\mathrm{e}} \) (Fig.~\ref{fig:7}(b)), suggesting that our result is reliable.

We also evaluate $ \zeta $ by measuring the diffusion coefficient along flow direction \cite{BobbiliMilner-31, IannirubertoMarrucci-27}. It is seen that the result is consistent with the $ \zeta $ found by the chain dragging method at a semi-quantitative level.

\textbf{\begin{figure}[H]
		\centering
		\includegraphics[width=\linewidth]{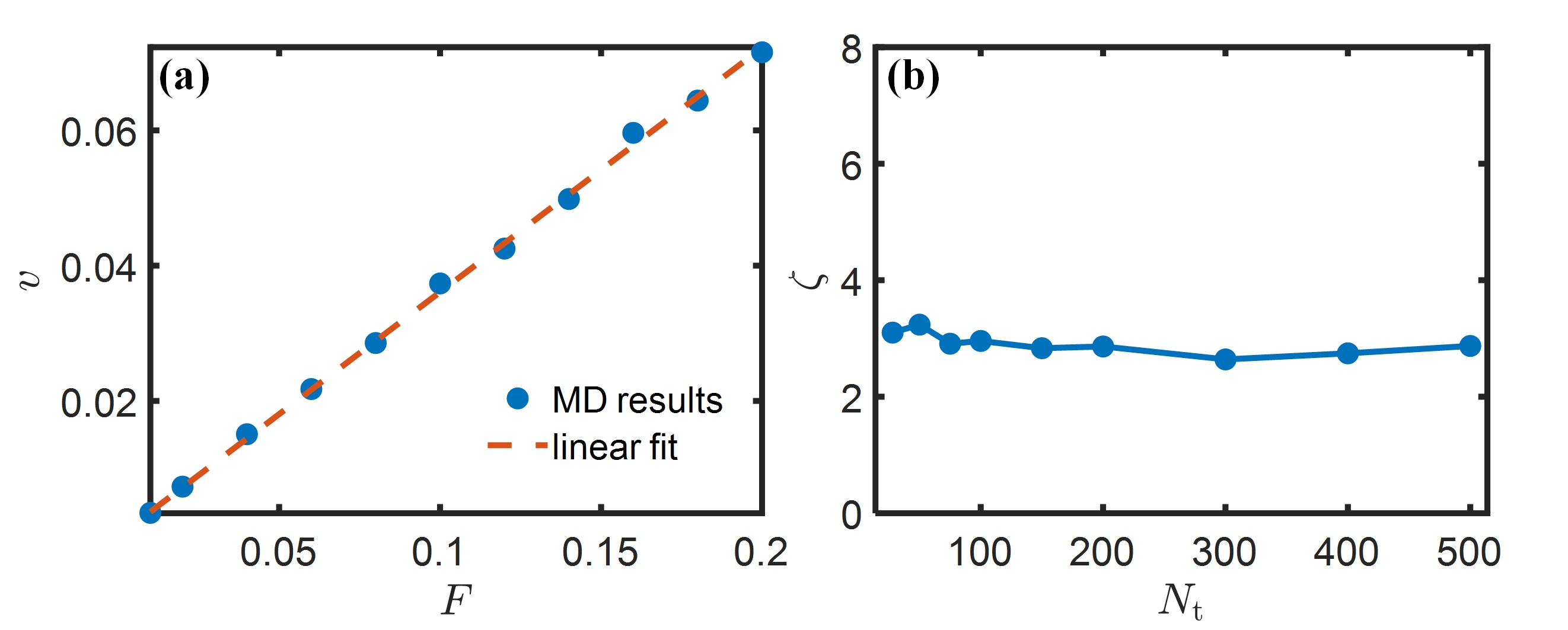}
		\caption{Chain-dragging results for the sample with \( N = 500 \), \( k_{\mathrm{bc}} = 1.5 \) and \( \mathrm{Wi}_{\mathrm{R}} = 40 \) at \( \varepsilon_{\mathrm{H}} = 6.0 \). (a) Applied force \( F \) vs center-of-mass velocity \( v \) of a test chain with \( N_{\mathrm{t}} = 500 \); symbols: MD results, dashed line: fit of $ v=\frac{1}{\zeta}F $ (b) Extracted monomer friction coefficient \( \zeta \) as a function of \( N_{\mathrm{t}} \) at fixed \( F = 0.08 \), showing a plateau for \( N_{\mathrm{t}} \geq N_{\mathrm{e}} \).}
		\label{fig:7}
\end{figure}}

\end{document}